\documentclass[aps,prb,twocolumn,superscriptaddress,oneside,floatfix,amsmath,showpacs,amssymb,footinbib]{revtex4-1}
\usepackage{graphicx}
\usepackage{dcolumn}
\usepackage{bm}

\begin{document}

\newcommand{\Y}{YBa$_2$Cu$_3$O$_{y}$~}
\newcommand{\ie}{\textit{i.e.}}
\newcommand{\eg}{\textit{e.g.}}
\newcommand{\etal}{\textit{et al.}}

%%%%%%%%%%%%%%%%%%%%%%%%%%%% TITLE

\title{Coherent \emph{c}-axis transport in the underdoped cuprate superconductor \Y}

%%%%%%%%%%%%%%%%%%%%%%%%%%%% AUTHORS
\author{B.~Vignolle}
\affiliation{Laboratoire National des Champs Magn\'{e}tiques
Intenses, UPR 3228, (CNRS-INSA-UJF-UPS), Toulouse 31400, France}

\author{B.~J.~Ramshaw}
\affiliation{Department of Physics and Astronomy, University of
British Columbia, Vancouver V6T 1Z1, Canada}

\author{James~Day}
\affiliation{Department of Physics and Astronomy, University of
British Columbia, Vancouver V6T 1Z1, Canada}

\author{David~LeBoeuf}
\affiliation{Laboratoire National des Champs Magn\'{e}tiques
Intenses, UPR 3228, (CNRS-INSA-UJF-UPS), Toulouse 31400, France}

\author{St\'{e}phane~Lepault}
\affiliation{Laboratoire National des Champs Magn\'{e}tiques
Intenses, UPR 3228, (CNRS-INSA-UJF-UPS), Toulouse 31400, France}

\author{Ruixing~Liang}
\affiliation{Department of Physics and Astronomy, University of
British Columbia, Vancouver V6T 1Z1, Canada} \affiliation{Canadian
Institute for Advanced Research, Toronto M5G 1Z8, Canada}

\author{W.N.~Hardy}
\affiliation{Department of Physics and Astronomy, University of
British Columbia, Vancouver V6T 1Z1, Canada} \affiliation{Canadian
Institute for Advanced Research, Toronto M5G 1Z8, Canada}

\author{D.A.~Bonn}
\affiliation{Department of Physics and Astronomy, University of
British Columbia, Vancouver V6T 1Z1, Canada} \affiliation{Canadian
Institute for Advanced Research, Toronto M5G 1Z8, Canada}

\author{Louis~Taillefer}
\affiliation{Canadian Institute for Advanced Research, Toronto M5G
1Z8, Canada} \affiliation{D\'epartement de physique \& RQMP,
Universit\'e de Sherbrooke, Sherbrooke, Qu\'{e}bec J1K 2R1,
Canada}

\author{Cyril~Proust} \email{cyril.proust@lncmi.cnrs.fr}
\affiliation{Laboratoire National des Champs Magn\'{e}tiques
Intenses, UPR 3228, (CNRS-INSA-UJF-UPS), Toulouse 31400, France}
\affiliation{Canadian Institute for Advanced Research, Toronto M5G
1Z8, Canada}

\date{\today}

%%%%%%%%%%%%%%%%%%%%%%%%%%%% ABSTRACT

\begin{abstract}

The electrical resistivity $\rho_c$ of the underdoped cuprate
superconductor \Y was measured perpendicular to the CuO$_2$ planes
on ultra-high quality single crystals in magnetic fields large
enough to suppress superconductivity. The incoherent
insulating-like behavior of $\rho_c$ at high temperature,
characteristic of all underdoped cuprates, is found to cross over
to a coherent regime of metallic behavior at low temperature.
This crossover coincides with the emergence of the small electron
pocket detected in the Fermi surface of \Y via quantum
oscillations, the Hall and Seebeck coefficients and with the
detection of a unidirectional modulation of the charge density as
seen by high-field NMR measurements.
The low coherence temperature is quantitatively consistent with
the small hopping integral $t_{\perp}$ inferred from the splitting
of the quantum oscillation frequencies.
We conclude that the Fermi-surface reconstruction in \Y at dopings
from $p = 0.08$ to at least $p = 0.15$, attributed to stripe
order, produces a metallic state with 3D coherence deep in the
underdoped regime.

\end{abstract}


\maketitle

%%%%%%%%%%%%%%%%%%%%%%%%%%%% INTRODUCTION
\section{Introduction}
Hole-doped cuprate superconductors stand out because of the presence of
an enigmatic pseudogap phase in the underdoped regime.\cite{Timusk99} From ARPES
measurements, the pseudogap is defined by the lack of a
well-defined quasiparticle peak in the anti-nodal region of the Brillouin zone leading
to Fermi arcs.\cite{Norman98} Among the peculiar properties in the pseudogap phase, the dichotomy between the
insulating-like inter-plane resistivity along the \emph{c}-axis and the metallic in-plane resistivity down to $T_c$ in
many underdoped cuprates is still heavily debated.\cite{Cooper94, Clarke97, Hussey07}
This dichotomy has been considered as strong support of the spin-singlet approach to the pseudogap phase,\cite{Anderson92} which consider that coherent charge transport is strictly bidimensional and \emph{c}-axis resistivity should diverge as $T \to 0$. Among other models that have been proposed, one of these is based on incoherent tunneling between layers assisted by interplanar disorder (see ref. \onlinecite{Gutman07} and references therein). Finally, recent valence-bond dynamical mean-field calculations have established a clear connection between the peculiar \emph{c}-axis charge transport and the lack of coherent quasiparticles as the pseudogap opens in the antinodal region.\cite{Ferrero10} This model is in good agreement with \emph{c}-axis optical conductivity measurements which reveal the absence of a Drude peak at low frequencies in the underdoped normal state above $T_c$.\cite{Basov05}\\
To understand \emph{c}-axis transport, not only must the Fermi arcs observed by ARPES above $T_c$ be taken into account but also the observation of quantum oscillations showing that the Fermi surface (FS) of underdoped cuprates undergoes a profound
transformation at low temperature.\cite{Doiron07} Combined with the negative Hall
\cite{LeBoeuf07} and Seebeck coefficients \cite{Chang10, Laliberte11} at low
temperature, these measurements demonstrate that the FS of
underdoped \Y is made of small electron pockets in contrast to the
large, hole-like FS of the overdoped cuprates.\cite{Vignolle08}
The underlying translational symmetry breaking, which causes the
Fermi surface reconstruction, has been observed in high fields nuclear magnetic resonance (NMR)
measurements.\cite{Wu11}
These microscopic measurements have revealed that at low temperature and above a threshold magnetic field, the translational symmetry of the CuO$_2$ planes in \Y is broken by the emergence of a unidirectional modulation of the charge density. This conclusion is supported by a comparative study of thermoelectric transport in underdoped \Y and in La$_{1.8-x}$Eu$_{0.2}$Sr$_x$CuO$_4$  - a cuprate where stripe order is well-established from X-ray diffraction \cite{Fink11} - which argues in favor of a charge stripe order causing reconstruction of the FS at low temperature for $p>0.08$.\cite{Laliberte11} This charge stripe order can naturally be interpreted as a competing order
with superconductivity akin to the smectic stripe phase observed
in the archetypal La$_{1.6-x}$Nd$_{0.4}$Sr$_x$CuO$_4$.\cite{Tranquada95}\\

\begin{figure*}
\centering
\includegraphics[width=0.95\linewidth,angle=0,clip]{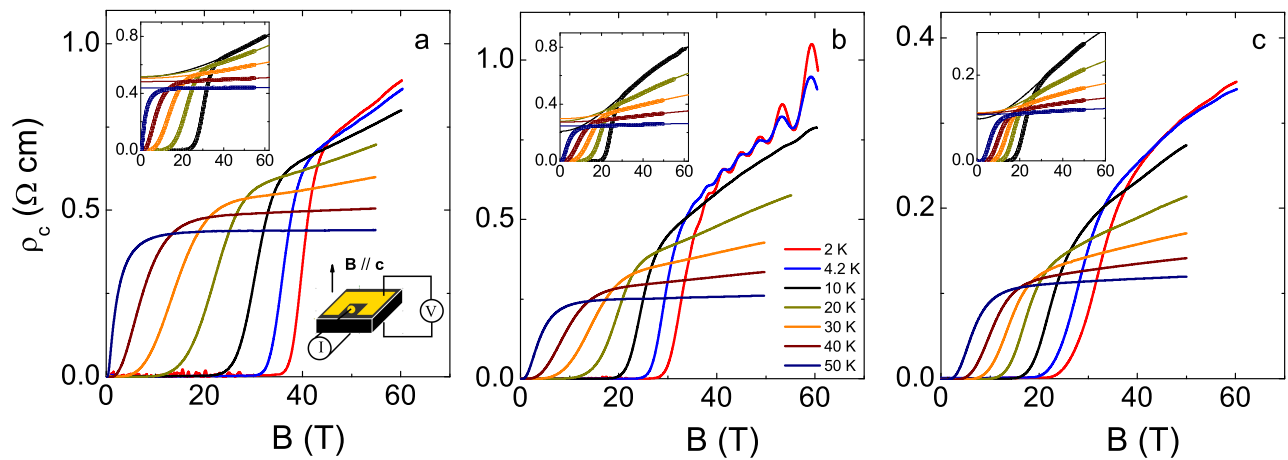}
\caption{ Electrical resistivity  $\rho_c$ of \Y
for a current \emph{I} and a magnetic field \emph{B} along the
\emph{c} axis ($I \parallel B \parallel c$). Three underdoped samples
were measured at different temperatures below $T_c$ (as indicated)
in pulsed magnetic fields up to 60 T. The doping level of each sample
is: (a) $p$ = 0.097, (b) $p$ = 0.109, and (c) p = 0.120.
Insets: Same data between 10 and 50~K
with a fit of each isotherm (thin solid lines) using a two-band model
above the superconducting transition (see section V).}\label{MRvsB}
\end{figure*}

Here we directly address the consequence of the Fermi surface reconstruction by stripe-charge order on \emph{c}-axis charge transport by measuring the \emph{c}-axis resistivity at low temperature in magnetic fields large enough to suppress superconductivity.
We found that the \emph{c}-axis resistivity becomes metallic-like
at low temperature, and interpret this as a consequence of
\emph{c}-axis coherence. The low coherence temperature implies a
small \emph{c}-axis dispersion and therefore a small hopping integral $t_{\perp}$,
in quantitative agreement with the splitting of the multiple
quantum oscillation frequencies.\cite{Audouard09, Sebastian10, Ramshaw11, Sebastian12} The
onset of this crossover coincides with the FS reconstruction leading to the emergence of a high mobility electron pocket which produces metallic-like transport both in the plane and along the \emph{c} axis. The coherence temperature decreases as the doping level decreases and vanishes at a hole
doping $p\approx0.08$, corresponding to the doping level where the
electron pocket disappears.\cite{LeBoeuf11}

\section{Sample preparation and experimental setup}

The samples studied were single crystals of YBa$_2$Cu$_3$O$_{y}$, grown in
non-reactive BaZrO$_3$ crucibles from high-purity starting
materials and subsequently detwinned.\cite{Liang00} The superconducting transition temperatures have been obtained by resistivity measurements at zero field: $T_c$ = 57.0~K ($p$ = 0.097), $T_c$ = 61.3~K ($p$ = 0.109) and $T_c$ = 66.4~K ($p$ = 0.120).  The doping $p$ of each
crystal was inferred from its superconducting transition
temperature $T_c$.\cite{Liang06} Electrical contacts to the
sample were made by evaporating gold, with large current pads and
small voltage pads mounted across the top and bottom so as to
short out any in-plane current (see inset of Fig.~\ref{MRvsB}a). Several samples with typical dimensions ($1 \times 1 \times t$) mm$^3$
of different thicknesses \emph{t} = 0.05 - 0.15~mm were measured, each giving similar values
of the absolute \emph{c}-axis resistivity. The resistivity was
measured at the LNCMI in Toulouse, France, in pulsed magnetic fields up to
60~T. A current excitation of 5~mA at $\approx$ 60~kHz was used.
The voltage (and a reference signal) was digitized using a
high-speed digitizer and post-analysed to perform the phase
comparison.

\section{c-axis magnetoresistance}

\begin{figure}
\centering
\includegraphics[width=0.95\linewidth,angle=0,clip]{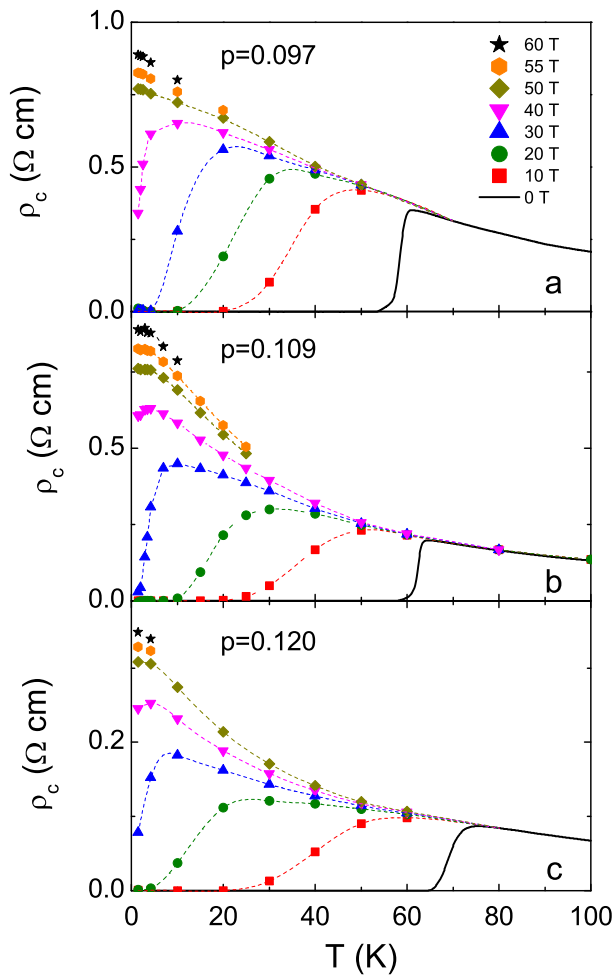}
\caption{ Electrical resistivity $\rho_c$ of \Y: (a) $p = 0.097$, (b)
$p = 0.109$ and (c) $p = 0.120$ plotted as a function of temperature
for different values of the magnetic field. Dashed lines are a
guide to the eye. The increase of the in-field \emph{c}-axis resistivity
down to about 4~K is in part due to the strong magnetoresistance
which develops at temperatures below 60~K.}\label{RT}
\end{figure}

Figure~\ref{MRvsB} presents the longitudinal \emph{c}-axis
resistivity up to 60 T for the three underdoped samples of YBa$_2$Cu$_3$O$_{y}$. Below 60~K, a strong positive magnetoresistance (MR)
grows with decreasing temperature, in good agreement with earlier
high-field measurements \cite{Balakirev00} of $\rho_c$ on \Y
crystals with $T_c=60$~K. At very low temperature, quantum oscillations are most clearly seen in the sample with $p$ = 0.109 (see Fig.~\ref{MRvsB}(b)) and are just above the noise level for the other two (Fig.~\ref{MRvsB}(a) and \ref{MRvsB}(c)). They arise from the quantization of cyclotron orbits perpendicular to the magnetic field. The frequencies and temperature dependence of these oscillations are consistent with previous reports.\cite{Doiron07,Audouard09,Sebastian10,Ramshaw11} Two features
common to all three samples are the rise of the in-field \emph{c}-axis
resistivity down to about 4~K, and a tendency to saturation at lower
temperature. This behavior is best captured in Fig.~\ref{RT}
where the resistivity is plotted as a function of temperature, at different magnetic fields. As \emph{T} is lowered, the \emph{c}-axis resistivity first exhibits insulating behaviour, but instead of diverging as \emph{T} $\rightarrow$0, it crosses over to a regime where it tends to saturate at the lowest temperatures in all three samples.

\begin{figure}
\centering
\includegraphics[width=0.95\linewidth,angle=0,clip]{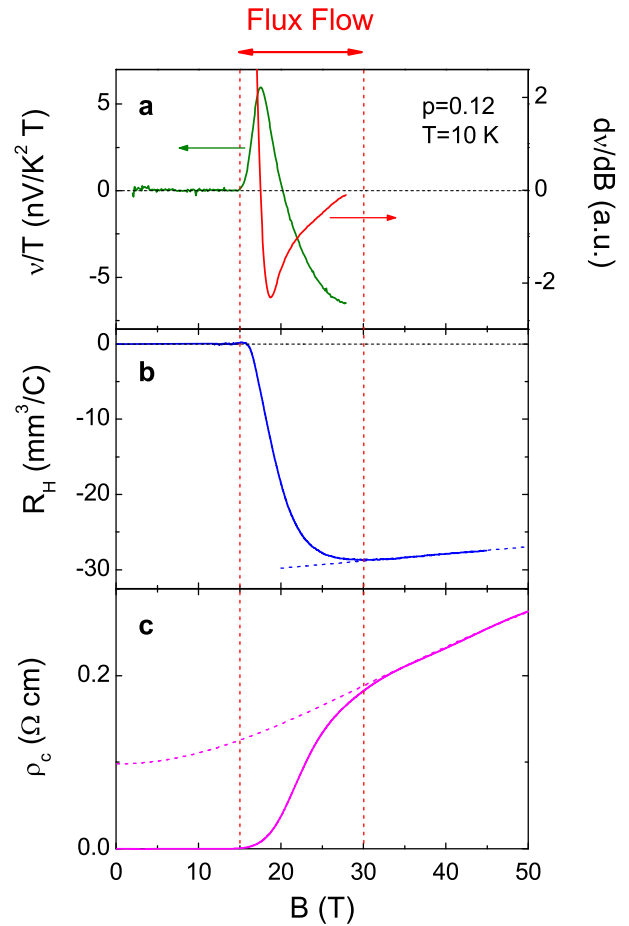}
\caption{ Field dependence of (a) the Nernst coefficient
\cite{Chang10}, (b) the Hall coefficient \cite{LeBoeuf07} and (c)
the \emph{c}-axis resistivity (Fig. 1c) of YBa$_2$Cu$_3$O$_{y}$ at $p=0.12$ and
$T=10$~K. Above a threshold field of about 30~T (indicated by the
right vertical dashed red line), the Nernst coefficient $\nu$
(panel a; green curve) saturates to its negative quasiparticle
value, as demonstrated by its derivative (panel a; red curve)
which goes to zero as $B \approx 30$~T. This saturation shows that
the positive contribution to the Nernst coefficient from
superconducting fluctuations has become negligible above 30 T.
Above this field, the Hall coefficient is almost flat (dashed blue
line in panel b) and the two-band model fit to the \emph{c}-axis
resistivity (dashed magenta line in panel c) merges with the data.
We conclude that at 10~K and above 30~T the flux-flow contribution
to the transport properties is negligible, and the large
magnetoresistance at high field is purely a property of the normal
state at this doping level.}\label{NS}
\end{figure}

\section{Flux flow contribution to the resistivity}

Upon cooling, the resistivity drops because of superconductivity.
This drop starts at lower temperature for higher fields. We define
the threshold field beyond which the normal state is reached as
the field above which  $\rho_c(T)$ shows no drop. To confirm that the tendency to saturation of  $\rho_c(T)$ at low temperature is not due to
some compensation between superconducting drop (as seen for the data below 40~T) and insulating-like
normal-state resistivity, we show that the tendency to saturation persists at
fields above a threshold field down to the
lowest temperatures (see Fig.~\ref{RT}). In all
three samples, 50~T is above the threshold field down to the
lowest temperatures. Therefore the magnetoresistance measured at
50~T is purely a normal-state property, at all three dopings.\\
The Nernst effect is a sensitive probe of flux flow, because
moving vortices make a large positive contribution to the Nernst
coefficient. In Fig.~\ref{NS}, we compare the field dependence
of the Nernst coefficient measured in \Y at $p=0.12$ and $T=10$~K \cite{Chang10} (Fig.~\ref{NS}a) with that of the in-plane Hall coefficient \cite{LeBoeuf07} (Fig.~\ref{NS}b) and
\emph{c}-axis resistivity (Fig.~\ref{NS}c). The Nernst
coefficient develops a strong positive peak above the melting line
due to vortex motion in the vortex liquid phase and is followed by
a gradual descent to negative values (the quasiparticle
contribution) until it becomes almost flat as the field approaches
30~T. This saturation is best captured by the field derivative of
the Nernst coefficient shown by a red line in Fig.~\ref{NS}a.
Above the threshold field of about 30~T, the Hall coefficient
becomes flat (as indicated by the blue dashed line in
Fig.~\ref{NS}b) and the two-band model used to fit the normal
state \emph{c}-axis resistivity merges with the data (see
Fig.~\ref{NS}c and Sec. V). This comparison confirms that flux-flow
contribution to the normal-state transport is negligible, that is to say that the magnetoresistance is entirely due to
quasiparticles, for fields greater than 30~T at $T$=10~K for \emph{p} = 0.120.

\begin{figure}
\centering
\includegraphics[width=0.95\linewidth,angle=0,clip]{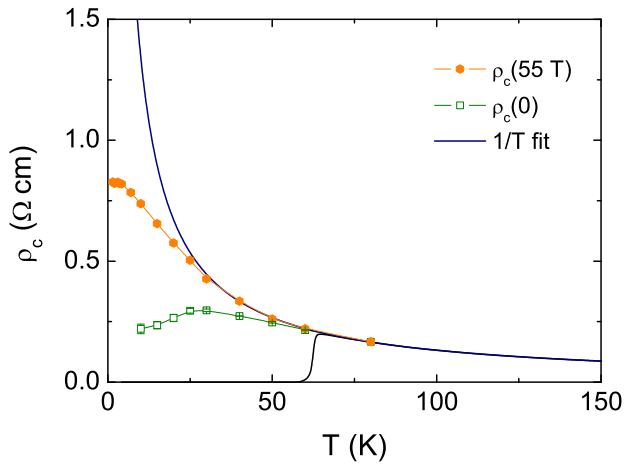}
\caption{ Temperature dependence of the
\emph{c}-axis resistivity of \Y (\emph{p}=0.109) measured at zero
magnetic field (black solid line) and at $B=55$~T (orange symbols).
Blue solid line is a $1/T$ fit to the high temperature data up to
300~K. The saturation of the in-field resistivity at low
temperature contrasts with the insulating-like behavior seen at
high temperature. The green open squares correspond to the
resistivity from which the magnetoresistance has been subtracted
using a two-band model to extrapolate the normal-state data of
Fig. 1 to \emph{B} = 0 (see section V).}\label{fig4}
\end{figure}

\noindent The same conclusion can be drawn for the other samples thanks to Hall effect measurements.\cite{LeBoeuf07} The key observation is that $T_0(B)$, the temperature at which $R_H(T)$ changes sign, is independent of field at high fields (from 40~T up to 60~T) for samples in the doping range studied here.
This shows that the temperature-induced sign change in $R_H$ at high
fields is not caused by flux flow and is thus clearly a property
of the normal state.\\ Earlier high-field measurements of
$\rho_c$ in \Y \cite{Balakirev00} show a striking difference
between the large magnetoresistance observed in samples with
$T_c=60$~K and the absence of the magnetoresistance in sample with
$T_c=49$~K. This can only be due to normal state transport
properties and can be explained by the vanishing of the very
mobile electron pocket for the low doping sample.\cite{LeBoeuf11}
There is no alternative explanation in terms of flux flow.

\section{Crossover towards coherent c-axis transport}

In fields of 50~T and above, where the \emph{c}-axis resistivity tends to saturate as
$T \rightarrow 0$, the non-superconducting ground state of \Y is
coherent in all three directions at $p=0.10 - 0.12$. This behavior is best captured in Fig.~\ref{fig4}, where the resistivities
for the sample with $p$ = 0.109 measured at zero field (solid black line) and at
$B=55$~T (orange symbols) are compared with the insulating-like high temperature behavior, where the \emph{c}-axis resistivity diverges as $1/T$ (blue solid line).\\
The observation of \emph{c}-axis coherent transport means that coherent Bloch bands along the \emph{c}-axis are present at low temperature and that charge carriers are not confined to the CuO$_2$ planes.\cite{Clarke97} Compared to the situation in overdoped cuprate superconductors, where there is no doubt about the existence of a three-dimensional Fermi surface,\cite{Hussey03} the \emph{c}-axis coherent transport in underdoped \Y appears at low temperature where the FS is reconstructed. NMR measurements performed on the same sample (\emph{p}=0.108) and in the same temperature/field range where quantum oscillations have been observed,\cite{Wu11} reveal an unidirectional charge-stripe order below a temperature $T_{charge}=50\pm10$~K above a threshold field $B \approx 20$~T. Not surprisingly, the transition temperature $T_{charge}$ obtained from NMR measurement coincides roughly with the temperature $T_0$, the temperature at which $R_H(T)$ changes sign.

\begin{figure}
\centering
\includegraphics[width=0.95\linewidth,angle=0,clip]{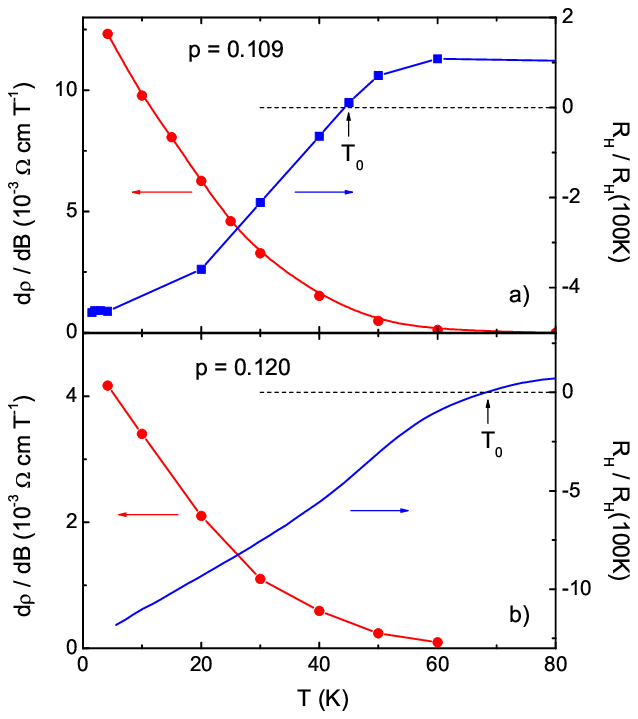}
\caption{ The slope of the \emph{c}-axis
magnetoresistance evaluated at $B=50$~T (red circles; left axis)
and in-plane Hall coefficient $R_H$ (blue squares; right axis) of
a) \Y (\emph{p}=0.109), b) \Y (\emph{p}=0.120) as a function of temperature. $R_H$ data measured at
a) $B=54$~T , and b) $B=45$~T  is taken from Ref.~\onlinecite{LeBoeuf11} and normalized
by its value at $T=100$~K. $T_0$ is the temperature at which
$R_H(T)$ changes sign from positive at high temperature to
negative at low temperature \cite{LeBoeuf07,LeBoeuf11}.}\label{MR-RHvsT}
\end{figure}

In the magnetic field / temperature range where the FS undergoes a reconstruction driven by the charge-stripe order, a high mobility electron pocket dominates the transport properties.
Although recent specific heat \cite{Riggs11} measurements performed at high fields point to a
Fermi surface made of only one pocket per CuO$_2$ plane,\cite{Harrison11} the
emergence of a strong non $B^2$ MR at low temperature (see Fig.~\ref{MRvsB}) is naturally explained
by the FS reconstruction into electron and hole sheets, due to the
ambipolar character of the Fermi surface. In Fig.~\ref{MR-RHvsT} we
compare the slope of the magnetoresistance $\rho_c(B)$ and the
in-plane Hall coefficient $R_H$ as a function of temperature
measured at the same hole doping.  The onset of the MR in $\rho_c$
coincides with the FS reconstruction thus revealing the two roles
played by the electron pocket: it enhances the orbital MR due to
in-plane motion of carriers and it allows the MR to be
reflected in inter-plane transport.

\begin{figure}
\centering
\includegraphics[width=0.95\linewidth,angle=0,clip]{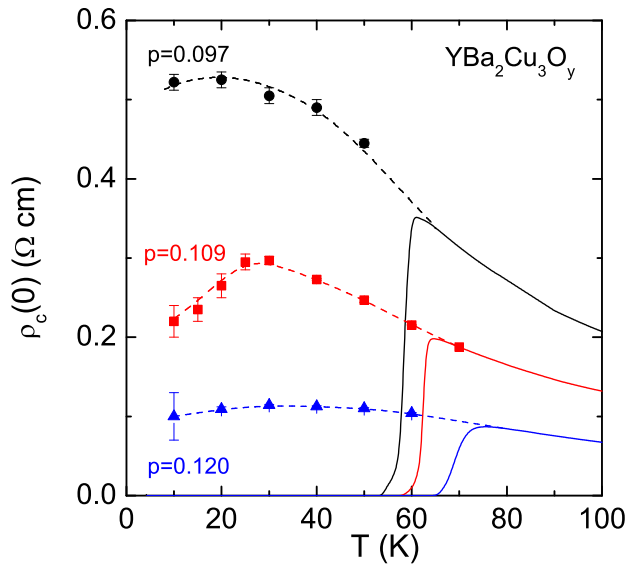}
\caption{ Temperature dependence of the \emph{c}-axis resistivity
of \Y from which the magnetoresistance has been subtracted using a
two-band model (see section V) for the three samples, as labeled. Solid lines show the resistivity measured in zero magnetic field. Dashed lines are a guide to the eye.}\label{RTfit}
\end{figure}

To reveal that $\rho_c$ is metallic-like at low temperature, it is necessary to obtain the MR-free temperature
dependence of $\rho_c(T)$ by extrapolating the in-field
resistivity $\rho_c(B)$ to $B=0$, defined as $\rho_c(0)$. Since a
strong MR develops at low temperature, any smooth extrapolation to
$B=0$ will give the same trend for the temperature dependence of
$\rho_c(0)$, namely an initial rise with decreasing temperature
turning into a drop at low temperature. To illustrate this, we
extrapolate the in-field resistivity $\rho_c(B)$ to $B=0$ using
the same two-band model (electron and hole carriers) that
self-consistently accounted for the temperature and field
dependence of the longitudinal and transverse (Hall) resistivities
of YBa$_2$Cu$_4$O$_8$.\cite{Rourke10} The transverse magnetoresistance can be
fitted with a two-band model:

%\begin{equation}
\begin{eqnarray} \label{MR}
\rho(B) & = &\frac{(\sigma_h+\sigma_e)+\sigma_h \sigma_e (\sigma_hR_h^2+\sigma_eR_e^2)B^2}{(\sigma_h+\sigma_e)^2+\sigma_h^2\sigma_e^2(R_h+R_e)^2B^2} \nonumber \\
\rho(B) & = &\rho_0+\frac{\alpha B^2}{1+\beta B^2}
\end{eqnarray}
%\end{equation}

where $\sigma_h$ ($\sigma_e$) is the conductivity of holes (electrons)
and $R_h$ ($R_e$) is the Hall coefficient for hole (electron)
carriers. Using the three free parameters $\rho_0$, $\alpha$ and
$\beta$, we were able to subtract the orbital magnetoresistance
from the field sweeps and get the temperature dependence of the
zero-field resistivity $\rho_c(0)=\rho_0(T)$. To estimate
error bars, we fitted each field sweep data set to Eq.~\ref{MR}
between a lower bound $B_{cut-off}$ and the maximum field strength
and monitored the value of $\rho_c(0)$ as a function of
$B_{cut-off}$. The resulting fits to $\rho_c(B)$ down to 10~K for all three samples are shown in thin solid lines in
the insets of Fig.~\ref{MRvsB}. They yield the extrapolated
zero-field resistivity $\rho_c(0)$ shown in symbols in
Fig.~\ref{RTfit}. The initial rise in $\rho_c(0)$ with decreasing
temperature turns into a drop at low temperature, passing through
a maximum at $T_{coh}=27 \pm 3$~K for the $p$ = 0.109 sample. This is in reasonable agreement
with the energy scale $t_{\perp} \approx 15$~K obtained from the
splitting of frequencies in quantum oscillations for \Y at \emph{p} =
0.10 - 0.11.\cite{Audouard09,Ramshaw11} The same analysis for the
two other compositions yields $T_{coh}=20 \pm 5$~K for $p$ = 0.097 and
$T_{coh}=35 \pm 5$~K for $p$ = 0.12.

\section{Discussion}

\begin{figure}
\centering
\includegraphics[width=0.95\linewidth,angle=0,clip]{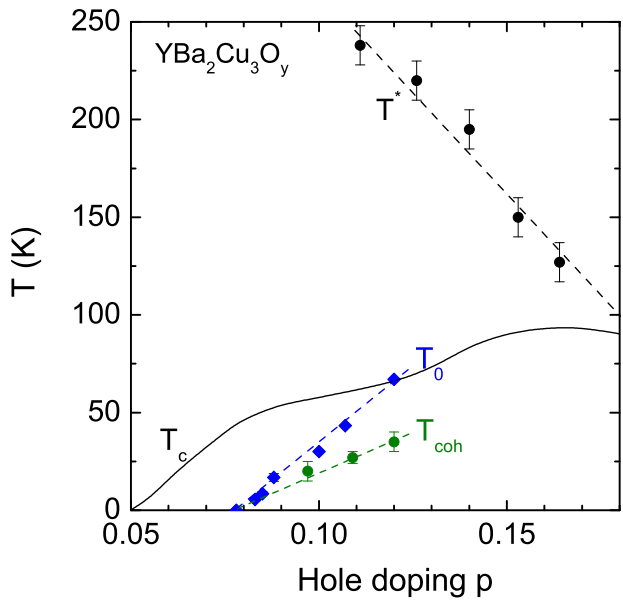}
\caption{ Temperature-doping phase diagram of YBa$_2$Cu$_3$O$_{y}$,
with the superconducting phase in zero magnetic field delineated
by the transition temperature $T_c$. Black circles mark the
temperature $T^*$ below which the in-plane resistivity deviates
from its linear temperature dependence at high temperature (from
data in Ref.~\onlinecite{Ando04}), a standard definition for the
onset of the pseudogap phase. The coherence temperature $T_{coh}$ (green circles) is the temperature where $\rho_c(T)$
peaks (see Fig.~\ref{RTfit}).The
coherence crossover splits the phase diagram into two regions: an
incoherent 2D regime above and a coherent 3D regime below. $T_0$
is the temperature at which the normal-state in-plane Hall
coefficient $R_H(T)$ of \Y changes sign from positive at high
temperature to negative at low temperature (blue diamonds; from
Ref.~\onlinecite{LeBoeuf11})}\label{PD}
\end{figure}

The coherent \emph{c}-axis transport at low temperature is a direct consequence of the Fermi
surface reconstruction occurring at a temperature scale
$T_{charge} \approx T_0$. From \emph{c}-axis transport
measurements, we define $T_{coh}$ as the characteristic temperature for
the crossover to the coherent regime at which $\rho_c(0)$ peaks
(see Fig.~\ref{RTfit}). In Fig.~\ref{PD}, we compare the
temperatures $T_{coh}$ and $T_0$ as a function of doping on the phase diagram of \Y.
The two phenomena trend similarly as a function of doping,
both decreasing to lower \emph{T} with decreasing \emph{p}. In addition, these
characteristic temperatures extrapolate to zero at lower doping $p
\approx 0.08$, where $R_H(T)$ no longer shows any downturn (see
data for sample with $T_c$ = 44.5 K in
Ref.~\onlinecite{LeBoeuf11}). This qualitative change has been
attributed to the disappearance of the electron pocket, either
caused by a Lifshitz transition \cite{LeBoeuf11} or by a phase
transition.\cite{Sebastian10b} Earlier measurements of
\emph{c}-axis transport on a sample with $T_c=49$~K
\cite{Balakirev00} are consistent with such a transition: the MR
in $\rho_c(T)$ is entirely gone and $\rho_c$ is now incoherent,
increasing down to the lowest temperatures.

In the framework of conventional theory of metals, the \emph{c}-axis
conductivity is given by $\sigma_c=\frac{4 e^2 c t_{\perp}^2 m^*
\tau_c}{\pi \hbar^4}$ where \emph{c} is the \emph{c}-axis lattice
parameter, $\tau_c$ is the relaxation time, and $m^*$ is the
effective mass. For a tetragonal cuprate material and due to the crystallographic structure,\cite{Andersen95} the interlayer
hopping integral $t_{\perp}$ depends strongly on the in-plane
momentum \textbf{k} of carriers, namely:\cite{Chakravarty93}
$t_{\perp}(\textbf{k})=\frac{t_{\perp}^0}{4}[cos(k_xa)-cos(k_yb)]^2$.
It is maximum at the anti-node i.e., at the ($\pi$, 0) (and
equivalent) points in the Brillouin zone. Between the pseudogap temperature $T^*$ and the temperature characteristic of the FS reconstruction $T_0$, the insulating-like behavior in $\rho_c$ and the absence of the Drude peak in \emph{c}-axis optical conductivity measurements in the underdoped regime \cite{Basov05} has been ascribed to the absence of a well-defined quasiparticle peak in the anti-nodal regions as seen by ARPES measurements.\cite{Norman98}\\
Below $T_0$ and for doping levels $p > 0.08$, the field-induced charge stripe order detected by high-field NMR in YBa$_2$Cu$_3$O$_{y}$ \cite{Wu11} causes the reconstruction of the FS at low temperature. The role of the magnetic field is to weaken superconductivity to reveal the competing stripe order. As a consequence, there is no contradiction between the low temperature / high field electronic properties of underdoped \Y and the FS seen by ARPES at zero field \cite{Norman98} or the absence of a Drude peak at low frequencies at magnetic fields below the threshold field.\cite{Tranquada10} This FS reconstruction produces a high mobility electron pocket which dominates the in-plane transport properties \cite{LeBoeuf11} and is responsible for the \emph{c}-axis coherence at low temperature.\\
Taking into account the in-plane momentum dependence of the interlayer hopping integral $t_{\perp}$, the cross-over to the coherent \emph{c}-axis transport suggests that either some electronic states exist close to the anti-node in the Brillouin zone \cite{Millis07,Chakravarty08,Yao11} or that the reconstructed FS allows for assisted interlayer hopping term through 1D bands for example. In the former scenario, strong scattering at (0, $\pi$) associated with charge fluctuations at a wavevector $Q_x$=[$\pi$/2a, 0] softens or freezes out when the charge order sets in at low temperature, restoring coherent quasiparticles close to the anti-node.
Below a doping level $p < 0.08$, the electron pocket disappears, probably because of a transition from a charge-stripe order ($p>0.08$) to a phase with spin order ($p<0.08$).\cite{Haug10} In the absence of this electron pocket, both in-plane \cite{LeBoeuf11} and out-of-plane \cite{Balakirev00} transport properties are non-metallic-like at low temperature.

%%%%%%% ACKNOWLEDGEMENTS  %%%%%%%%%%%%%%%%

We thank A. Carrington, S. Chakravarty, A. Chubukov, M.-H. Julien,
S. Kivelson, A. Millis, M. Norman, R. Ramazashvili, T. Senthil, G.
Rikken and M. Vojta for useful discussions. Research support was
provided by the French ANR DELICE, Euromagnet II, the Canadian
Institute for Advanced Research and the Natural Science and
Engineering Research Council.

\end{document}